\documentclass[12pt,preprint]{aastex}
\usepackage{amsmath}

\newcommand{\Swift}{{\it Swift}}
\newcommand{\Suzaku}{{\it Suzaku}}
\newcommand{\Swifts}{{\it Swift }}

\slugcomment{ApJ Letter (accepted)}
\shorttitle{Optical afterglow jet break with known Epeak}
\shortauthors{Urata  et al.}

\begin{document}

\title{\Swifts GRB GRB071010B : outlier of the $\rm E^{src}_{peak} - E_{\gamma}$ and ${\rm E_{iso}-E^{src}_{peak}-t^{src}_{jet}}$ correlations }

\author{
Yuji~\textsc{Urata}\altaffilmark{1,2,3}, 
Kuiyun~\textsc{Huang}\altaffilmark{3},
Myungshin~\textsc{Im}\altaffilmark{4}
Induk~\textsc{Lee}\altaffilmark{4,1}
Jinsong~\textsc{Deng}\altaffilmark{5},
WingHuen~\textsc{Ip}\altaffilmark{1},
Hans~\textsc{Krimm}\altaffilmark{6,7},
Xin~\textsc{Liping}\altaffilmark{5},
Masanori~\textsc{Ohno}\altaffilmark{8},
Yulei~\textsc{Qiu}\altaffilmark{5},
Satoshi~\textsc{Sugita}\altaffilmark{9,10},
Makoto~\textsc{Tashiro}\altaffilmark{2},
Jianyan~\textsc{Wei}\altaffilmark{5},
Kazutaka~\textsc{Yamaoka}\altaffilmark{10},
and 
Weikang~\textsc{Zheng}\altaffilmark{5}
}

\altaffiltext{1}{Institute of Astronomy, National Central University, Chung-Li 32054, Taiwan; urata@astro.ncu.edu.tw}
\altaffiltext{2}{Department of Physics, Saitama University, Shimo-Okubo, Saitama, 338-8570, Japan}
\altaffiltext{3}{Academia Sinica Institute of Astronomy and Astrophysics, Taipei 106, Taiwan}
\altaffiltext{4}{Center for the Exploration of the Origin of the Universe, Department of Physics \& Astronomy, FPRD, Seoul National University, Shillim-dong, San 56-1, Kwanak-gu, Seoul, Korea}
\altaffiltext{5}{National  Astronomical Observatories, Chinese Academy of Sciences, Beijing 100012, China}
\altaffiltext{6}
{NASA Goddard Space Flight Center, Greenbelt, MD 20771, USA}
\altaffiltext{7}
{Universities Space Research 
Association, 10211 Wincopin Circle, Suite 500, Columbia, MD 21044, USA}
%
%
\altaffiltext{8}{Institute of Space and Astronautical Science, Japan Aerospace Exploration Agency, 3-1-1 Yoshinodai, Sagamihara, Kanagawa 229-8510, Japan}
\altaffiltext{9}{The Institute of Physical and Chemical Research (RIKEN), 2-1 Hirosawa, Wako, Saitama, 351-0198, Japan}
\altaffiltext{10}{Department of Physics and Mathematics, Aoyama Gakuin University, 5-10-1, Fuchinobe, Sayamihara 229-8558, Japan}
%

\begin{abstract}

  We present multi-band results for GRB071010B based on \Swift,
  \Suzaku, and ground-based optical observations. This burst is an
  ideal target to evaluate the robustness of the ${\rm
    E^{src}_{peak}-E_{iso}}$ and ${\rm E^{src}_{peak}-E_{\gamma}}$
  relations, whose studies have been in stagnation due to the lack of
  the combined estimation of $\rm E^{src}_{peak}$ and long term
  optical monitoring. The joint prompt spectral fitting using
  \Swift/Burst Alert Telescope and \Suzaku/Wide-band All sky Monitor
  data yielded the spectral peak energy as E$^{src}_{peak}$ of
  $86.5^{+6.4}_{-6.3}$ keV and E$_{iso}$ of
  $2.25^{+0.19}_{-0.16}\times10^{52}$ erg with $z=0.947$. The optical
  afterglow light curve is well fitted by a simple power law with
  temporal index $\alpha=-0.60\pm0.02$. The lower limit of temporal
  break in the optical light curve is 9.8 days. Our multi-wavelength
  analysis reveals that GRB071010B follows ${\rm
    E^{src}_{peak}-E_{iso}}$ but violates the ${\rm
    E^{src}_{peak}-E_{\gamma}}$ and ${\rm
    E_{iso}-E^{src}_{peak}-t^{src}_{jet}}$ at more than the 3$\sigma$
  level.

\end{abstract}

\keywords{gamma rays: bursts --- gamma rays: observation}

\section{Introduction}


A few tight correlations linking several properties of gamma-ray
bursts (GRBs), namely the spectral peak energy, the total radiated
energy, and the afterglow break time, have been discovered with
pre-\Swifts GRBs.  \citet{amati} found the empirical relation between
the rest-frame spectral peak energy of prompt emission ${\rm
  E^{src}_{peak}}$ (here, we use the E$^{src}_{peak}$,
E$^{obs}_{peak}$ for the values in the rest and observer frames,
respectively) and the isotropic-equivalent energy released during the
prompt phase ${\rm E_{iso}}$. \citet{g04} showed a tight empirical
relation between the ${\rm E^{src}_{peak}}$ and the
collimation-corrected energy ${\rm E_{\gamma}}$ based on a jet
interpretation of the temporal break in the optical afterglow light
curve.
Liang \& Zhang (2005) found the three-dimensional correlation in the ${\rm E_{iso}-E^{src}_{peak}-t^{src}_{jet}}$ plane.
The existence of these correlations could uncover crucial properties
of GRB physics which are not yet fully understood.
Recently, there have been studies of outliers to these relations
(e.g., Campana et al 2007, Sato et al. 2007, Ghirlanda et al. 2007).
Using \Swift/XRT observations, \citet{sato} showed that there are cases
with continuous X-ray afterglow light curves where no jet break is
seen at the expected break time derived from the ${\rm
  E^{src}_{peak}-E_{\gamma}}$ relation. However, some X-ray afterglow
light curves may not show jet breaks together with the optical
afterglow at the expected time, because there is a possibility that
X-ray and optical emission come from different regions and mechanisms
even in the normal decay phase (e.g., Urata et al. 2007a, Liang et al
2007, Huang et al. 2007, Ghisellini et al. 2009).  Therefore, it is
critical to check the aforementioned relations by combining prompt
$\gamma$-ray wide-band spectroscopy and optical long-term monitoring.


In this Letter, we present joint spectral analysis of the prompt
gamma-ray and systematic optical follow-up results for GRB 071010B.
This event was detected and localized by \Swift \citep{gyro}. The
\Suzaku/Wide-band All sky Monitor (WAM; Yamaoka et al. 2009) and
Konus/{\it WIND} \citep{konus} also detected this main burst.  Thanks
to the wide energy band of the \Suzaku/WAM (Yamaoka et al. 2009), the
joint spectral fittings between the \Swift/Burst Alert Telescope (BAT)
and the \Suzaku/WAM data yield better constraint for the determination
of $E_{\rm peak}$.
The optical afterglow was discovered by \citet{6873} and observed by
many telescopes from the early stage (e.g., Wang et al. 2007). The
redshift was determined to be $z=0.947$ by \citet{redshift} using the
Gemini North telescope.  In the frame work of EAFON (Urata et
al. 2003, 2005), we have performed long term monitoring to determine
the jet break time.
Therefore, GRB071010B is one of the most suitable event to evaluate
${\rm E^{src}_{peak}-E_{iso}}$, ${\rm E^{src}_{peak}-E_{\gamma}}$
and ${\rm E_{iso}-E^{src}_{peak}-t^{src}_{jet}}$ in \Swifts era.
All the errors are quoted at the 90\% statistical confidence level in this
paper.

\section{Observations}

\subsection{\Swift/BAT}

\begin{figure}
\epsscale{.80}
\plotone{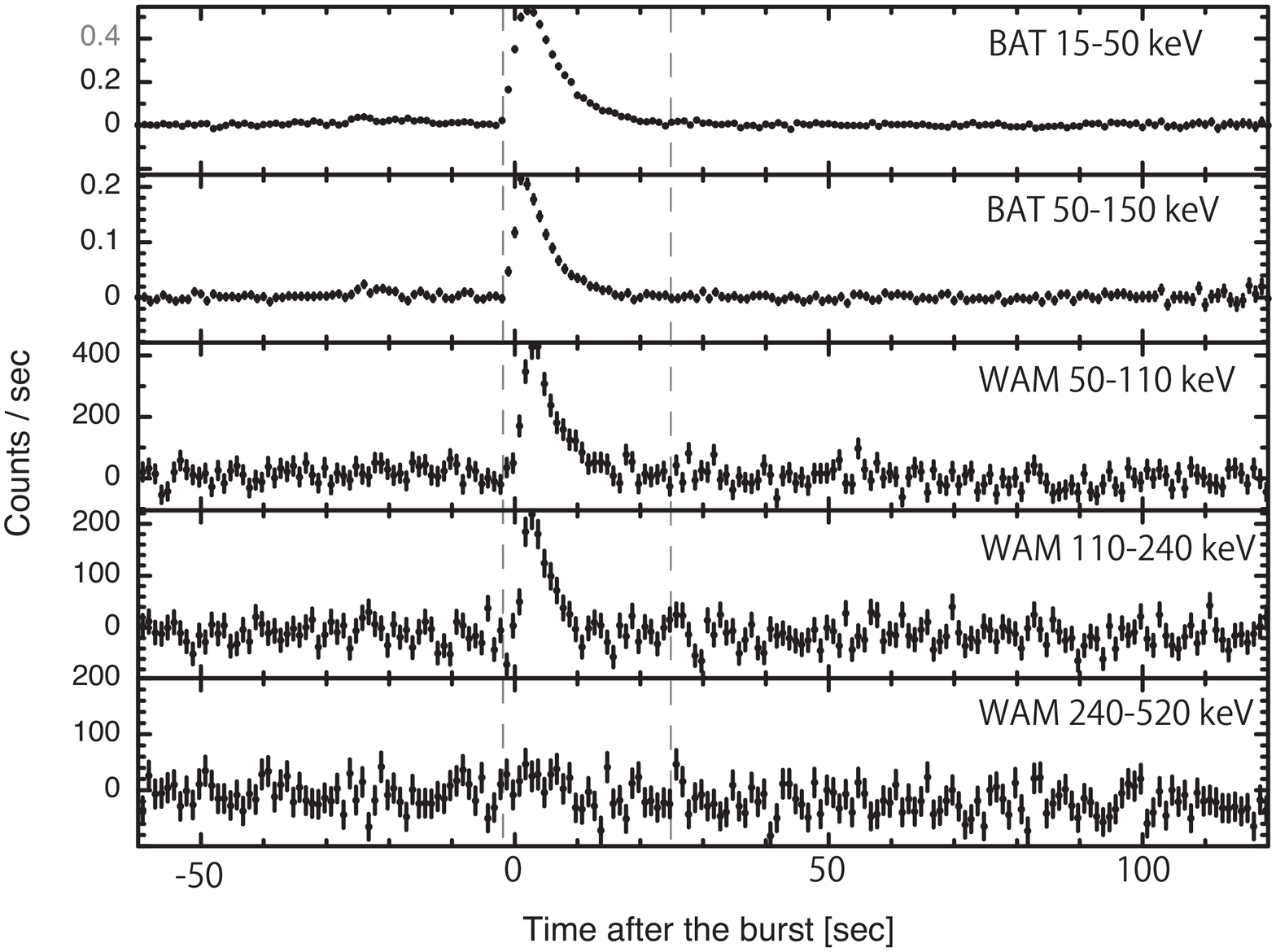}
\caption{Prompt X-ray and gamma-ray light curve of GRB071010B
  observed by \Swift/BAT and \Suzaku/WAM. The data in the time region between the
  dashed lines are used for the joint spectral analysis.\label{promptlc}}
\end{figure}

The \Swifts/BAT triggered and located GRB
071010B (trigger ID=293795) at 20:45:47 (T$_{0}$) UT on 2007 October
10.  The on-board localization was distributed 13 sec after the
trigger and its position was reported as RA$=10^{\rm h}02^{\rm
  m}07^{\rm}.5$, Dec$=+45^{\circ}44'0"$, with an uncertainty of
$3'$. The spacecraft did not slew promptly to the burst position
because automated slewing was disabled due to ongoing recovery from a
gyro anomaly \citep{gyro}. After 6800 s of the trigger, \Swift/XRT
started observing the field and detected a bright X-ray counterpart
at R.A.$=10^{\rm h}02^{\rm m}09^{\rm}.2$, decl.$=+45^{\circ}43'52".2$,
with an uncertainty of $5"$.
\citet{bat} reported the mask-weighted light curve which shows a
pre-trigger pulse starting at T$_{0}\sim-45$ s, peaking at
T$_{0}\sim-20$ s, and returning almost to background at T$_{0}\sim-8$
s. The main FRED pulse started at T$_{0}\sim2$ s and ended around
T$_{0}\sim60$ s. A small third soft peak started at T$_{0}+95$ s and
was terminated by the planned spacecraft slew \citep{bat}.
The two peaks at T$_{0}-45$ s and T$_{0}+95$ s are much weaker
than the main peak, thus they are not visible in Figure \ref{promptlc}.

\subsection{\Suzaku/WAM}

The {\it Suzaku}/WAM was triggered at 20:45:49 (T$_{0}+2$ s) UT on
2007 October 10. The \Suzaku/WAM is the active shield of the Hard
X-ray detector \citep{hxd,hxd2} onboard the 5th Japanese X-ray
satellite {\it Suzaku} \citep{suzaku}. It consists of large area,
thick BGO crystals and is also designed to monitor the all sky from 50
keV to 5 MeV by a large effective area. The large effective area from
300 keV to 5 MeV ($400{\rm cm^2}$) surpasses those of other currently
operating experiments with spectral capability, and enables us to
perform wide-band spectroscopy of GRBs with high sensitivity.

As shown in Figure \ref{promptlc}, the WAM light curve also shows a
single FRED-like peak starting at T$_{0}-1$ s, ending at T$_{0}+8$
sec.  The duration of the WAM signal is $T90=5$ s.  This spike was
also detected by both the \Swift/BAT, and the Konus/{\it WIND}. There
is no significant signal from the pre-cursor and the weak soft tail
seen in the \Swift/BAT light curve \citep{bat} during the time
coverage.

\subsection{Optical Follow-ups}

We carried out follow-up observations of the GRB071010B optical
afterglow at the Mt. Lemmnon, Arizona, USA and the Xinglong, China
observatories within the framework of the EAFON (Urata et al. 2003,
2005).
Using the robotic 1 m telescope and a $2k\times2k$ CCD camera at the
Mt. Lemmon observatory operated by the Korea Astronomy Space Science
Institute (I. Lee et al. in preparation,; Han et al. 2005), we have made $B$-,
$V$-, and $R$-band imaging observations with 300 s exposures,
starting at 10:51:14.6 UT on 2007 October 11 (0.5871 days after the
burst). 
In all the bands, we detected the afterglow clearly. We have also
monitored the afterglow on October 12, 16, 19, 21, 25, and 26 in
the $Rc$-band.
Additional $V$-band observations were performed using the EST 1.0 m
telescope at the Xinglong Observatory \citep{weikang08}, starting at
2007 October 11 (19:08 UT; 0.971 days after the burst). The filed of
view is $11'.4\times 11'.1$, and the pixel size is $0''.51$ square.

The $i'$- and $z'$-band observations were acquired with the 3.6-m CFHT
using MegaCam on 2008 October 26, and November 6 (approximately one year
after the burst), respectively. These deeper images show the host
galaxy with $i'=23.63\pm0.14$ AB mag, $z'=22.73\pm0.11$ AB mag at R.A.$=10^{\rm
  h}02^{\rm m}09^{\rm}.274$, decl.$=+45^{\circ}43'49".69$.

\section{Analysis and Results}

\subsection{\Swift/BAT Prompt Emission}

The BAT data were analyzed using the standard BAT analysis software
distributed within HEADAS v6.4. Using batgrbproduct v2.41, the
mask-weighted BAT light curves were created in the standard four energy
bands, 15 -- 25, 25 -- 50, 50 -- 100, 100 --350 keV, and the duration was
determined as T$_{90}=36.0\pm2.4$ s. Figure \ref{promptlc} shows
15 -- 50 keV and 50 -- 150 keV light curves.
Response matrices were generated with the task batdrmgen using the
latest spectral redistribution matrices.
The time-averaged spectrum (15 -- 150 keV) from T$_{0}-35.7$ to
T$_{0}+24.2$ s is well fitted by a power law with an exponential
cutoff. This fit yields a photon index of $1.53\pm0.22$, and
E$^{obs}_{peak}$ of $52.0\pm6.4$ keV.  A fit to a simple power law
gives a photon index of $2.01\pm0.05$ ($\chi^{2}/\nu=0.82$ for
$\nu=57$).

\subsection{\Suzaku/WAM Prompt Emission}

The WAM spectral and temporal data were extracted using hxdmkwamlc and
hxdmkwamspec in the HEADAS version 6.4.  The background was estimated
using the fitting model described in Sugita et al. (2009).  Response
matrices were generated by the WAM response generator as described in
\citet{ohno}. The time-averaged spectrum (150 -- 1000 keV) from
T$_{0}-0.78$ to T$_{0}+24.2$s was well fitted by a single power law with
a photon index of $2.64^{+0.26}_{-0.22}$ ($\chi^{2}/\nu=1.07$ for
$\nu=10$).
This result is consistent with the {\it Swift}/BAT power law with
an exponential cutoff fitting with the photon index $1.53\pm0.22$, and
E$^{obs}_{peak}$ of $52.0\pm6.4$ keV.

\subsection{BAT and WAM Joint Analysis}

\begin{figure}
\epsscale{.80}
\plotone{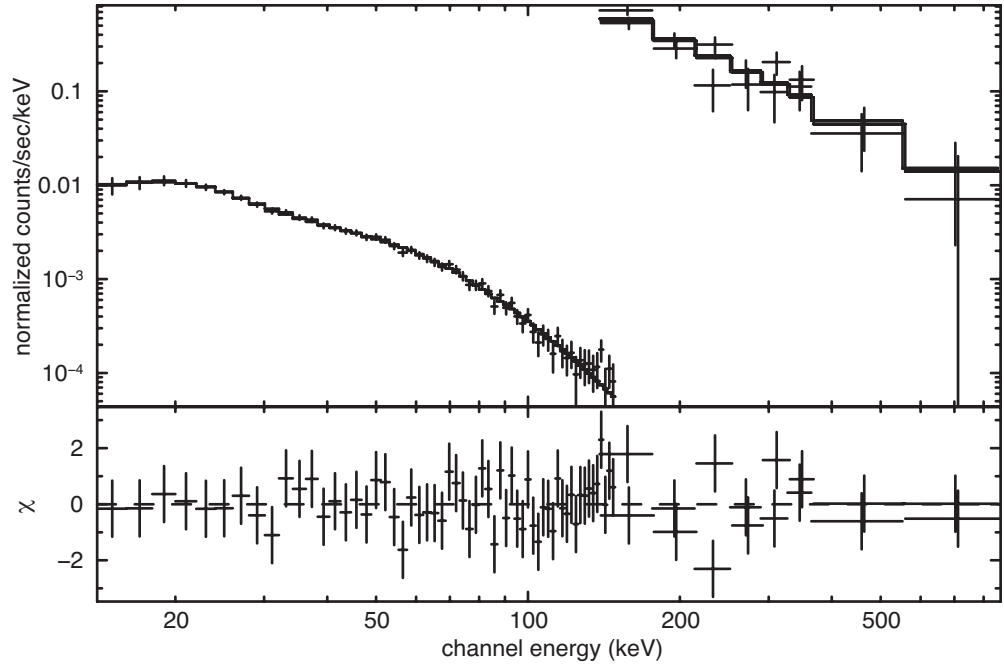}
\caption{Broad band spectrum of GRB071010B prompt emission made by the joint analysis of the \Swift/BAT and \Suzaku/WAM data.\label{spec}}
\end{figure}

In order to better constrain the spectral peak energy, we perform
joint fitting between {\it Swift}/BAT and {\it Suzaku}/WAM.  Figure
\ref{promptlc} shows the overlapping time regions (from T$_{0}-0.780$ to
T$_{0}+24.2$s) for this fitting. As shown in Figure \ref{spec}, the
spectrum is well fitted with the Band function (Band et al 1993). The
fitting yields a low-energy photon index of $1.19^{+0.29}_{-0.23}$,
a high-energy photon index of $2.30^{+0.06}_{-0.07}$ and
E$^{src}_{peak}$ of $86.5^{+6.4}_{-6.3}$ keV ($\chi^{2}/\nu=0.59$ for
$\nu=87$).
This result is consistent with that of  Konus/{\it WIND} \citep{konus}.
We have also estimated the $\rm E_{iso}$ as
$2.25^{+0.19}_{-0.16}\times 10^{52}$ erg, assuming cosmological
parameters: $H_0=71 {\rm km s^{-1} Mpc^{-1}}$, $\Omega_m=0.27$, and
$\Omega_{\Lambda}=0.73$.

\subsection{Optical Afterglow}

A standard routine including bias subtraction and flat-fielding
corrections was employed to process the data using the {\it IRAF}
package. The DAOPHOT package was used to perform aperture photometry
of the GRB images. For the photometric calibration of the afterglow,
several calibration stars reported by Henden (2007) were chosen.  In
Figure \ref{optlc}, we plot the $Rc$-band light curve of the
GRB071010B afterglow based on our photometry.  The data contain our
LOAO measurements, TAOS observations \citep{wang08} with the
re-calibrated GCN measurements (\citet{6873,6918,6923,6935,6919,6945}
). The light curve between 0.02 and 10.6 day is well fitted by a
single power law with index $\alpha = -0.60 \pm 0.02$ ($\chi^{2}/\nu$=
0.89 for $\nu=37$), defined by $F(t) \propto t^{\alpha}$, where $F(t)$
is the flux inthe $Rc$-band at time $t$ after the BAT trigger time
T$_{0}$. The single power-law index is consistent with that of the
TAOS measurement ($\alpha \sim -0.51$) reported by \citet{wang08}.
%
This fitting also excludes the possible 3.4 days temporal break reported by
Kann et al (2007) and Im et al (2007) based on GCN data point
measurements.
We have also confirmed that the host-galaxy brightness ($i'=23.63$ mag,
$z'=22.73$ mag) derived by the CFHT observations is insufficient to affect
this fitting. We consider both red and typical host-galaxy colors to
evaluate $R$-band brightness.
Although the late-time light curve could be fitted by a single
power-law, it is interesting to note that a possible temporal break
occurred at around 9 days after the burst. We tried to fit the
0.02--10.6 days light curve with a broken power law function expressed
as
\begin{equation}
 F_{\nu}(t) = {F_{\nu}^{\ast} \over [(t/t_b)^{\alpha_1}+(t/t_b)^{
 \alpha_2}]},
\end{equation}
where $\alpha_1$ and $\alpha_2$ are the power-law indices before and
after the break, $F_{\nu}^{\ast}$ is the flux at the break, and $t_b$ is the break
time.
The fitting yields $\alpha_{1}=-0.54\pm0.03$,
$\alpha_{2}=-3.03\pm1.15$ and $t_b=9.81\pm1.00$ ($\chi^{2}/\nu$=0.52
for $\nu=35$).
When we fix $\alpha_{2}$ as -2, the fitting provides the break time of $t_{b}=9.61\pm1.49$.
Therefore, these results imply that the temporal break should be
around or later than 9.8 days after the burst.

\section{Discussion}

Spectral parameters of the prompt emission of GRB071010B are well
constrained by our joint fitting of \Swift/BAT and \Suzaku WAM. The
measured values of current event (E$^{src}_{peak} =
86.5^{+6.4}_{-6.3}$ keV and $\rm E_{iso} = 2.25^{+0.19}_{-0.16}\times
10^{52}$ erg) well follow the ${\rm E^{src}_{peak}-E_{iso}}$
relation.  Furthermore, our systematic long term (10 days) optical
monitoring observation suggests a possible jet break at later than
$\sim 9.8$ days after the burst.
The wealth of the multi-wavelength data with a good temporal coverage
makes GRB071010B one of the best targets to evaluate the ${\rm
  E^{src}_{peak}-E_{\gamma}}$ and the ${\rm
  E_{iso}-E^{src}_{peak}-t^{src}_{jet}}$ relations which studies have
been in stagnation due to the lack of the $\rm E^{src}_{peak}$
estimation and long term optical monitoring.
%

\begin{figure}
\epsscale{.80}
\plotone{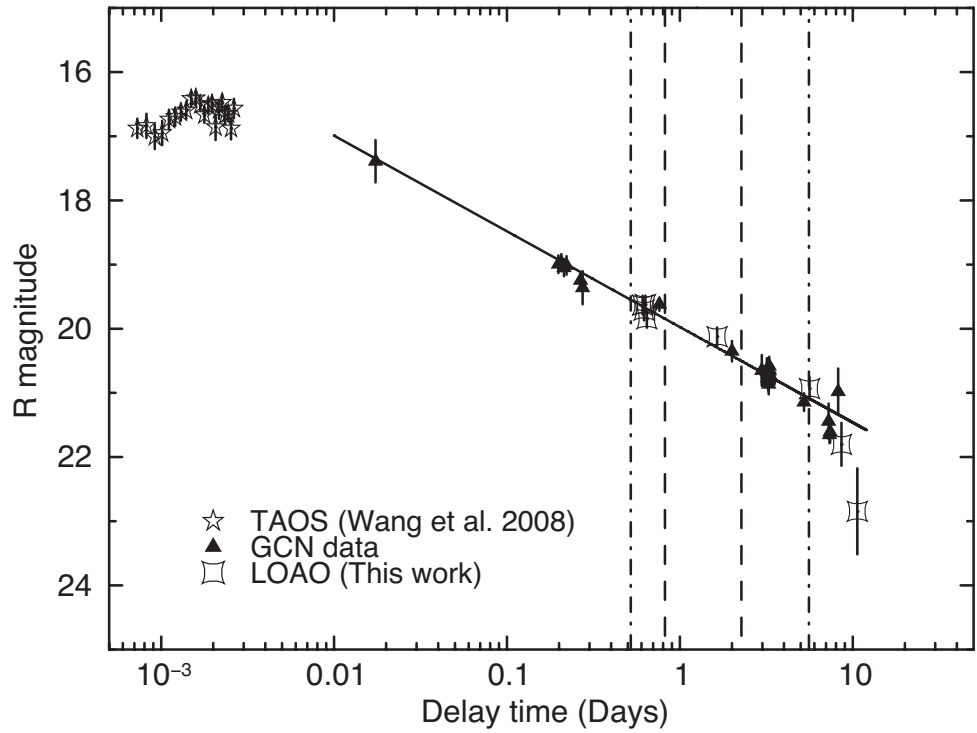}
\caption{$Rc$-band afterglow light curve. The time region enclosed by the dashed ($3\sigma$) and the dash-dotted ($5\sigma$) lines indicates the expected jet break time from the $\rm E^{src}_{peak} - E_{\gamma}$ relation.\label{optlc}}
\end{figure}

In order to evaluate the ${\rm E^{src}_{peak}-E_{\gamma}}$ relation for the
optical light curve, we invert this relation to predict the jet break
time as described in \citet{sato}. The expected
jet break time is expressed as
\begin{equation}
t_{jet} = 389\left(1+z\right)\left(\frac{n}{3 {\rm cm^{-3}}}\right)^{-\frac{1}{3}}\left(\frac{\eta_{\gamma}}{0.2}\right)^{-\frac{1}{3}}{\rm {E_{iso,52}}^{-1}}\left(\frac{{\rm E^{src}_{peak}}}{A}\right)^{1.89} {\rm day},
\end{equation}
where $n$ and $\eta_{\gamma}$ are the number density of the ambient
medium and the efficiency of the shock, respectively.  Here, we use
the Ghirlanda relation as $E^{src}_{peak}=A{E_{\gamma, 52}}^{0.70}$,
allowing $A$ to vary within the $3\sigma$ level.
The value of $n$ is in general within 1$<n<$30 cm$^{-2}$ (Panaitescu
\& Kumar 2001, 2002).
Following the assumption made by \citet{g04} for most of
their sample GRBs, we initially assume $n=3$ cm$^{-2}$ and
$\eta_{\gamma}$=0.2. 
%
%
As shown in Figure \ref{optlc}, the optical light curve is well
described with a single power law during the expected jet break time.
We also tried to fit the optical light curve by fixing the jet break
time as $t_{j}=1.25$ days expected from Equation (2). The fitting
yields $\alpha_{1}=-0.30\pm0.06$ and $\alpha_{2}=-0.91\pm0.04$
($\chi^{2}/\nu$=0.72 for $\nu=36$).  Although this fitting is
acceptable in terms of the $\chi^{2}$-fitting, an {\it F}-test indicates
that this broken power-law model with $t_j=1.25$ days over the single
power law is not significant at the 90\% confidence level.  
These temporal decay indices are inconsistent with those of typical
jet break events.
The temporal ($\alpha_x=0.67$) and spectral ($\beta_x=1.10$) relation
of X-ray afterglow also support the sphere phase ($\alpha=3/2\beta-1 \sim
0.65$). These X-ray values come from UK's X-ray afterglow repository
(Evans et al 2009).
Thus, these results imply that GRB071010B is an outlier of the tight
${\rm E^{src}_{peak}-E_{\gamma}}$ relation.
%



\begin{figure}
\epsscale{.80}
\plotone{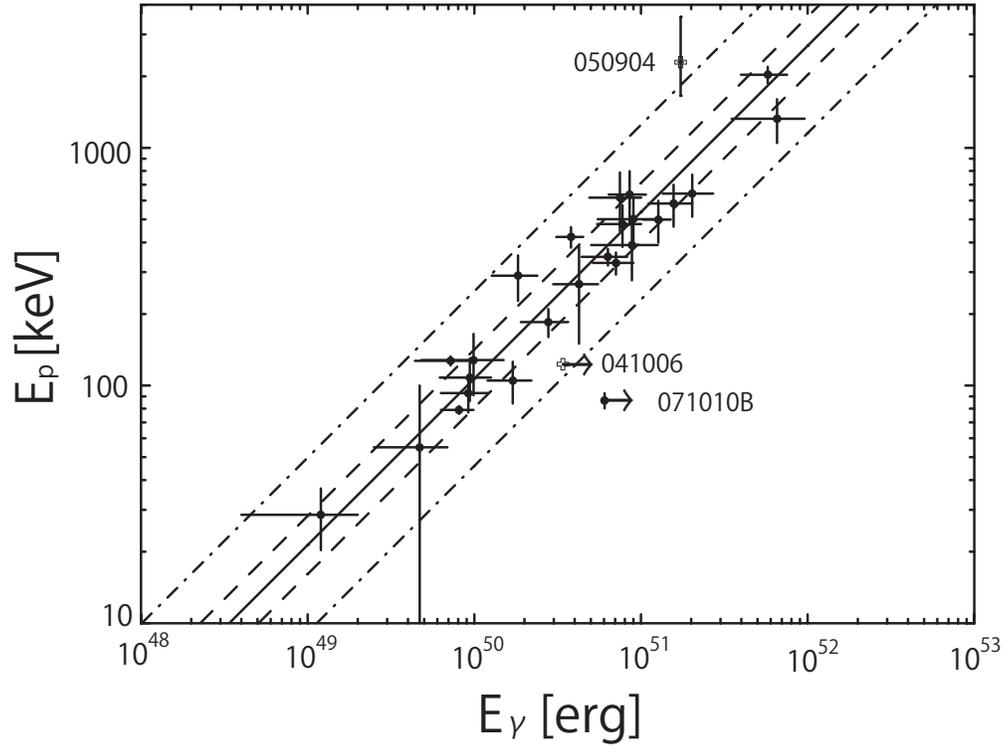}
\caption{$E^{src}_{peak}-E_{\gamma}$ relation including the data for GRB071010B and GRB041006 corrected for a homogeneous circumburst medium.
 The solid line shows the best-fit correlation
  derived by \citet{g07}. The dashed and dash-dotted lines indicate
  the 1$\sigma$ and 3$\sigma$ scatter of the correlation,
  respectively.\label{grel}}
\end{figure}

In Figure \ref{grel}, we also plotted the $\rm E^{src}_{peak} - E_{\gamma}$
relation together with GRB041006, GRB050904 and others reported by
\cite{g07}.  According to the detailed optical analysis of the
GRB041006 afterglow \citep{041006}, this pre-\Swifts era event shows a
chromatic optical plateau phase around $0.1\sim0.2$ days after the
burst. Since \citet{g07} regarded this chromatic plateau in the
GRB041006 optical light curve as the jet break, we updated $\rm
E_{\gamma}$ for this event.  As shown in Figure \ref{grel}, GRB071010B
is a clear outlier of the tight correlation with more than $3\sigma$
deviation.  GRB041006 is also possible outlier at the $2\sigma$ level.
We have also tested the wind case. The estimated E${_\gamma, _w} >
2.8\times 10^50$ erg from the jet break limit is inconsistent
with Ghirlanda et al (2007) by over $3\sigma$ level.
According to \citet{sugita}, GRB050904 is also an outlier of the $\rm
E^{src}_{peak} - E_{\gamma}$ relation, but it follows the ${\rm
  E^{src}_{peak}-E_{iso}}$ relation.


These three events including a high-redshift GRB050904 imply the ${\rm
  E^{src}_{peak}-E_{\gamma}}$ relation could have larger scatter than
originally suggested. Since the ${\rm E^{src}_{peak}-E_{\gamma}}$
correlation requires a small scatter to be used to standardize GRBs in
a way similar to Type Ia supernovae \citep{g04cosmo}, these three
events pose a problem to the cosmological application of the ${\rm
  E^{src}_{peak}-E_{\gamma}}$ relation.
Another implication is that these outliers have the relevant role of
the circumburst environments for different origins (e.g., long and short
bursts).
In the GRB050904 case, the gap can be explained by a higher
circumburst density which implies a massive stars origin.  The
required density is consistent with the estimations by optical and
radio observations \citep{sugita}.  However, GRB071010B requires a 2
or 3 order smaller circumburst density ($n\sim5.9 \times 10^{-3}
\rm{cm^{-3}}$) to be consistent with the tight correlation. This lower
density environment implies the origin of GRB071010B is of short GRB,
where typical circumburst density is $n=10^{-2} \rm{cm^{-3}}$
(e.g., Nakar 2007), even though the prompt duration of GRB071010B is
longer than 2 s (T$_{90}=36.0\pm2.3$ s).

\begin{figure}
\epsscale{.80}
\plotone{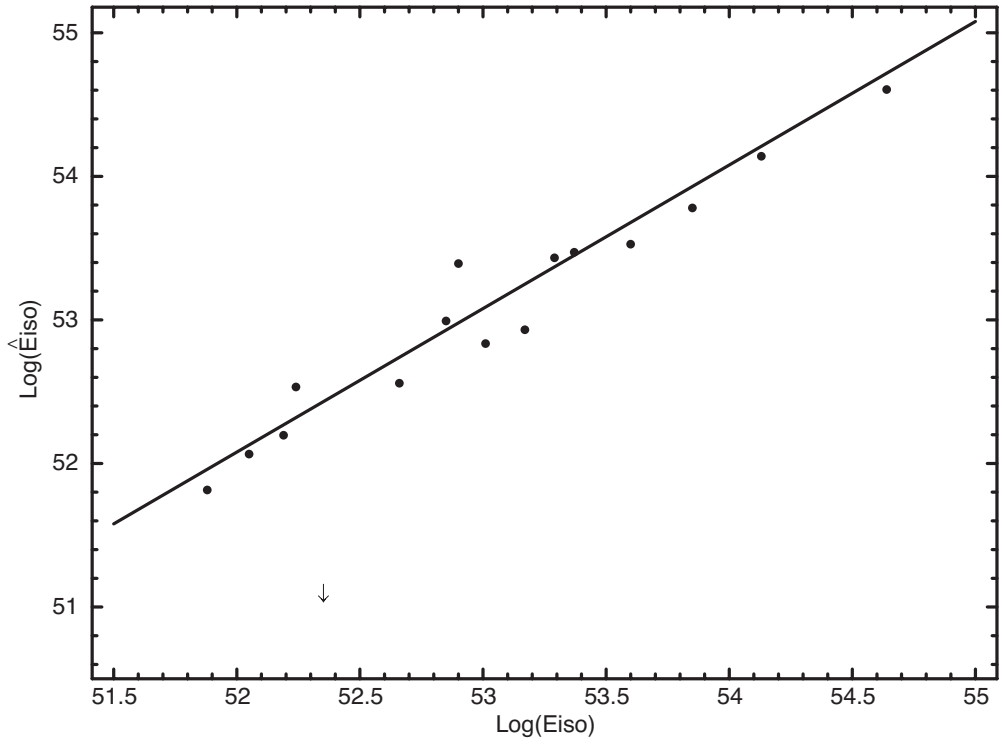}
\caption{Plot of log $\hat{E}_{\rm{iso}}$ calculated by the empirical ${\rm E_{iso}-E^{src}_{peak}-t^{src}_{jet}}$ relation \citep{liang05} as compared with log $E_{\rm{iso}}$ derived from the observed fluence. The arrow indicates the upper limit of GRB071010B.\label{liangzhang}}
\end{figure}

We also tested the empirical ${\rm E_{iso}-E^{src}_{peak}-t^{src}_{jet}}$
relation \citep{liang05} expressed as
\begin{equation}
{\rm E_{iso}}/10^{52} = (0.85\pm0.21)\left( \frac{E^{src}_{peak}}{100 {\rm keV}} \right)^{1.94\pm0.17} \left(\frac{t^{src}_{jet}}{1 {\rm day}} \right)^{-1.24\pm0.23},
\end{equation}
where $t^{src}_{jet}$ is the jet break time in the rest frame.
As shown in Figure \ref{liangzhang}, this relation yields the expected
upper limit of isotropic total energy as $\rm E_{iso} < 0.11 \times
10^{52} {\rm erg}$ with the jet break time $t_{jet}>9.8$
day. Therefore, GRB071010B is also an outlier of this
model-independent relation at more than the 3$\sigma$ level.  

Since the tight correlations $\rm E^{src}_{peak} - E_{\gamma}$ and
${\rm E_{iso}-E^{src}_{peak}-t^{src}_{jet}}$ might indicate a crucial
property of GRB physics (e.g., \citet{t06} and \citet{t07} for the $\rm
E^{src}_{peak} - E_{\gamma}$ relation), GRB071010B might be classified
as a new category.  However, there is also the possibility that these
relations simply have larger dispersion than previous studies.
Therefore, further study is required with large volume of good samples
which can be built up by the joint analysis of \Swift/BAT, \Suzaku/WAM,
Konus/{\it WIND}, Fermi/GBM and future missions such as the Space Variable
Objects Monitor (SVOM) \citep{svom} with ground-based optical
follow-ups.

\acknowledgments

We thank Bing Zhang for useful comments and
discussions.  M.I. and I.L. acknowledge the support by the Creative
Research Initiatives grant R16-2008-015-01000-0 of MOST/KOSEF.  This
work is partly supported by grants NSC 98-2112-M-008-003-MY3 (Y.U.), NSC
96WFA0700264 (W.H.I.), Ministryof Education under the Aim for Top
University Program NCU (W.H.I.), National Natural Science Foundation of
China (No. 10673014), and National Basic Research Program of China
(No. 2009CB824800).  Access to the CFHT was made possible by the
Ministry of Education and the National Science Council of Taiwan as
part of the Cosmology and Particle Astrophysics (CosPA) initiative.

\end{document}